
\input harvmac

\def\inbar{\,\vrule height1.5ex width.4pt depth0pt}
\def\IB{\relax{\rm I\kern-.18em B}}
\def\IC{\relax{\hbox{{$\inbar\kern-.3em{\rm C}$}}}}

\def\ID{\relax{\rm I\kern-.18em D}}
\def\IE{\relax{\rm I\kern-.18em E}}
\def\IF{\relax{\rm I\kern-.18em F}}
\def\IG{\relax\hbox{$\inbar\kern-.3em{\rm G}$}}
\def\IH{\relax{\rm I\kern-.18em H}}
\def\II{\relax{\rm I\kern-.18em I}}
\def\IK{\relax{\rm I\kern-.18em K}}
\def\IL{\relax{\rm I\kern-.18em L}}
\def\IM{\relax{\rm I\kern-.18em M}}
\def\IN{\relax{\rm I\kern-.18em N}}
\def\IO{\relax\hbox{$\inbar\kern-.3em{\rm O}$}}
\def\IP{\relax{\rm I\kern-.18em P}}
\def\IQ{\relax\hbox{$\inbar\kern-.3em{\rm Q}$}}
\def\IR{\relax{\rm I\kern-.18em R}}
\def\IZ{\relax\ifmmode\mathchoice
{\hbox{Z\kern-.4em Z}}{\hbox{Z\kern-.4em Z}}
{\lower.9pt\hbox{Z\kern-.4em Z}}
{\lower1.2pt\hbox{Z\kern-.4em Z}}\else{Z\kern-.4em Z}\fi}
\def\IGa{\relax\hbox{${\rm I}\kern-.18em\Gamma$}}
\def\IPi{\relax\hbox{${\rm I}\kern-.18em\Pi$}}
\def\ITh{\relax\hbox{$\inbar\kern-.3em\Theta$}}
\def\IOm{\relax\hbox{$\inbar\kern-3.00pt\Omega$}}

\def\half{{1 \over 2}}

\def\nub #1, #2, #3. { {\rm Nucl. Phys. }{\bf B#1} {\rm (#3) #2}}
\def\nubf #1, #2, #3, #4. { {\rm Nucl. Phys. }{\bf B#1 [FS#2]} {\rm (#4) #3}}
\def\plb #1, #2, #3.{ {\rm Phys. Lett. }{\bf #1B} {\rm (#3) #2}}
\def\lmpa #1, #2, #3. { {\rm Lett. Mod. Phys. }{\bf A#1} {\rm (#3) #2}}
\def\prl #1, #2, #3. { {\rm Phys. Rev. Lett. }{\bf#1} {\rm (#3) #2}}
\def\prd #1, #2, #3. { {\rm Phys. Rev. }{\bf D#1} {\rm (#3) #2}}
\def\pro #1, #2, #3. { {\rm Phys. Rev. }{\bf #1} {\rm (#3) #2}}
\def\jel #1, #2, #3. {{\rm JEPT Lett. }{\bf #1}{\rm (#3) #2}}
\def\jep #1, #2, #3. { {\rm Sov. Phys. JEPT }{\bf #1} {\rm (#3) #2}}
\def\cmp #1, #2, #3. { {\rm Comm. Math. Phys. }{\bf #1} {\rm (#3) #2}}
\def\half{{1 \over 2}}

\rightline{Technion-PH-23-91}
\Title{}{ String Winding in a Black Hole Geometry.}
\medskip
\centerline{{\bf Mordechai Spiegelglas }\footnote*{
Lady Davis Fellow at the Technion. ~~~~~~~~~{\tenpoint Bitnet:
phr74ms@technion.}}}
\smallskip
\centerline {Physics Department}
\centerline {Technion, Haifa 32000, Israel.}
\bigskip
\bigskip
\bigskip
\noindent
\vskip 2cm

\def\I{I}
\def\III{III}
\def\V{V}
\def\pfs{para\-fermions}

\def\pfic{para\-fermionic}

  {\abstractfont  $U(1)$ zero modes in the $SL(2,\IR)_k/U(1)$ and
$SU(2)_k/U(1)$ conformal coset theories, are investigated in
conjunction with the string black hole solution. The angular variable
in the Euclidean version, is found to have a double set of winding.
Region \III\  is shown to be $SU(2)_k/U(1)$ where the doubling
accounts for the cut sructure of the parafermionic amplitudes and fits
nicely across the horizon and singularity. The implications for string
thermodynamics and identical particles correlations are discussed.}

\vfill

\smallskip

\Date{July 1991}

  Recently, a particularly interesting string solution was found
\ref\jm{S. Elitzur, A. Forge and E. Rabinovici, \nub 359, 581, 1991.
.},  \ref\ind{G. Mandal, A.M. Sengupta and S.R. Wadia, \lmpa 6, 1685,
1991. .}, \ref\blw{E. Witten, {\it On String Theory and Black Holes},
IAS preprint  IASSNS-\-HEP/91/12 (1991).}. In \blw\ the solution for
the graviton-\-dilaton field equations, which shares many features
with a 3+1 dimensional black hole, was also found as the conformal
coset model $SL(2,\IR)/U(1)$. Duality, relating region \I\  ending at the
horizon to region \V\  starting at the singularity, became promptly a
particularly intriguing feature of this model \ref\amit{A. Giveon,
{\it Target Space Duality and Stringy Black Holes}, Berkeley preprint,
LBL-30671 (1991).}, \ref\dvv{R. Dijkgraaf, E. Verlinde and H.
Verlinde, {\it String Propagation in a Black Hole Geometry.}, IAS
preprint IASSNS-\-HEP/91/22 (1991)}. The later, also sets the formalism
and notation used here.

  We are going to discuss the $U(1)$ winding in this string solution.
In the Euclidean version, the line element for the cigar shaped region
I is given by  \eqn\gei{ds^2= dr^2 + \, \tanh ^2 r \, d \theta^2.} We
will find that $\theta$ has actually a double set of winding. This is
in contrast with the usual case in conformal field theory, where
right-moving modes and left-moving modes share the winding and the
primary operators those imply\foot{Doubling is thus related to ideas
presented in \ref\witorb{E. Witten, \prl 61, 670, 1988. .}\  about
orbifolds in complexified space-time and the separation of left and
right movers.}. Heuristically, the source for doubling is the
topological $U(1)/U(1)$ theory \ref\gigi{M. Spiegelglas and S.
Yankielowicz, {\it  $G/G$ - Topological Field Theories by Cosetting},
Tel-Aviv, Technion preprint, PH-34-90.}, \ref\wigi{E. Witten, {\it On
Holomorphic Factorization of WZW and Coset Models}, IASSNS-\-HEP/91/25
(1991).}\  hiding in the $SL(2,\IR)/U(1)$ coset construction as argued
at \ref\gak{K. Gaw{\c{e}}dzki and A. Kupiainen, \nub 320, 625, 1990.
.}, \ref\karb{ D. Karabali and  H.J. Schnitzer, \nub 329, 649, 1990.
.} (where the current-\-algebra $H \subset G$, was guaged to form
$G/H$ by complexifying $H$ and then employing complex BRST to cancel
the $H$ propagating degrees of freedom). The $U(1)/U(1)$ is
classically the theory of flat $U(1)$ gauge connections, or extra
winding for an angular variable $\theta$.

  Before elaborating on widing doubling, we would like to present a
somewhat broader context which makes flat gauge conections
particularly interesting around black hole \blw, a goemetry which may
have access to some non-perturbative aspects of string theory.
Independently of the coupling constant it has a horizon and a
singularity,  with gravitational interactions becoming strong at the
later. This non-perturbative flavor of the black hole solution is a
motivation to the search for string features on which it differs from
flat space-time. We will find that Euclidean winding is different
around the black hole than in the flat space-time case.

  Winding, has important roles in string theory. It is responsible for
extra gauge symmetries of string origin \ref\het{D. Gross, J. Harvey,
E. Martinec and R. Rohm,\prl 54, 502, 1985. , \nub 256, 253, 1985. ,
\nub 267, 75, 1986. .}. It is also important in string thermodynamics.
Strings in temperatue $1/\beta$, are formulated by curling one
(Euclidean space) dimension, to radius $\beta$ \ref\polch{J.
Polchinski, \cmp 104, 37, 1986. .}, \ref\atwi{J.J. Atick and E.
Witten, \nub 310,  291, 1988. .}. Increasing the temperture by
shrinking $\beta$, a string winding state, could have its $M^2 = -C +
t \,\beta^2$ becoming tachyonic (as the second winding term cannot
cancel the first tachyonic zero point term anymore). This was argued
to indicate a transition temperature for the string, namely Hagedorn
temperature, where the number of excited states wins over their
Boltzman supression and the free energy is dominated by highly excited
string states \ref\bala{B. Sathiapalan, \prd 35, 3277, 1987. .},
\ref\kog {Ya. I. Kogan, \jel 45, 709, 1987. .}.  The implications of
the winding doubling for string thermodynamics will be discussed
subsequently.

  An additional string aspect which seems different between the black
hole and flat space-time, is the way the Hilbert space is constrained
to a ghost-free positive normed physical spectrum \ref\shitau{ I am
grateful to S. Yankielowicz for raising this point.}. The well known
mechanism of BRST cohomology does not seem to help, at least in its
straight\-forward version, which requires a flat time-like direction
\ref\kaog{M. Kato and K. Ogawa, \nub 212, 443, 1983. .}, \ref\huch{M.
Spiegelglas, \nub 283, 205, 1987. , Talk at the XV Int. Coll. on Group
Theoretical Methods in Physics, Philadelphia, 1986, (World Scientific)
p. 639.}. There are suggestions, based on the $SL(2,\IR)_k$  current
algebra, giving negative-norm state free spectrum for the black-hole
string solution \ref\bars{I. Bars, {\it String Propagation on Black
Holes,} USC-91/HEP-B3 preprint.}. However, their compatibility with
modular invariance has yet to be settled, along with finding the
symmetry principle underlying them.

  A different way out is a non-linear physical state condition,
possibly an interacting string field equation like $Q \Psi + \Psi
\star \Psi=0$, for an appropriately defined string field theory (with
product $\star$), instead of the condition $Q \Psi =0 $ solved by BRST
cohomology. A non-linear condition of this kind could fit with the
non-perturbative nature of the black hole solution. Regretably, the
relevant version of string field theory is not clear yet. Nonetheless,
we could entertain the formal similarity between this physical state
condition and a flat gauge connection. Although flat gauge
configurations cannot be suggeted so far as a useful field theory
approximation to the relevant string field theory\foot{ In string
field theory $Q$ is constructed out of Virasoro generators $T(z)$ and
ghosts $b(z)$, $c(z)$. There is, however, the Sugawara formula gives
$T(z)$ as a bilinear of $SL(2,\IR)_k$ currents parallelled by $b(z)$
expressed as a product of $SL(2,\IR)_k$ ghost and a current \gigi.
With some luck, the similarity between the physical state condition,
the highest weight condition with respect to the $U(1)$ gauged in
$SL(2,\IR)_k$ and flat gauge condition may be more than formal.}, we
will proceed to study them.

  \def\i2{{i \over 2}}
  \def\thet{{\displaystyle \theta}}
  \def\sigm{{\displaystyle \sigma}}
  \def\zb{\overline{z}}

  Let us take stock of angular variables in the coset model $
SL(2,\IR)_k/U(1)$. It is convenient to specify the action of the
$U(1)$ gauge transformations in the WZW model \ref\wzw{E. Witten, \cmp
92, 455, 1984. } in the Euler notation for $g(z, \zb) \in SL(2,\IR)$
\eqn\eul{ g_{\lower1pt\hbox{$\scriptstyle \,{\rm I},~euc.$}}(z, \zb) =
e^{\i2 \thet_L \sigm_2} e^{\half r \, \sigm_1} e^{\i2 \thet_R
\sigm_2}. } The left handed (holomporhic) transformations act on
$\thet_L $ while the right handed (anti-holo\-mporhic) ones act on
$\thet_R$. The choice of the relative sign of the right handed gauge
action, gives  vectorial or axial $U(1)$ gauge transformations
\ref\krts{E.B. Kiritsis, {\it Duality in Gauged WZW Models}, LBL/UCB
preprint LBL-\-UCB-\-PTH-\-91/12} and its results are discussed below.
Anyway, gauging a $U(1)$ on the world-sheet will leave us with one
angular field, $\theta(z,\zb)$ and some doubts counting zero modes
associated with it. The Euclidean region \I\  can be found by gauging
the vectorial $U(1)$ generated by the Pauli matrix $\sigma_2$.  A
convenient gauge choice \blw\ is taking $g(z,\zb)$ symmetric, leading
to the line element \gei. One could try other gauge choices, like
taking $g(z,\zb)$  traceless, which yields a different angular
variable $\tilde{\theta}$ (and the line element $ds^2= dr^2 + \, \coth
^2 r \, d \tilde{\theta}^2$). In both cases the coset has a single
angular variable. $\theta$ is locally related to $\tilde {\theta}$ by
a $U(1)$ gauge transformation. If, however, the world-sheet is
topologically non-trivial, this would not mean that the two are
equivalent. It would rather mean that the ambiguity in the choice of
an angular field, is given by an extra $U(1)$ flat connection as a
global set of degrees of freedom.

  In our discussion we will often use space-time (or Euclidean space)
arguments for the black hole solution, which are usually justified
semi-classically at infinite level $k$ which is weak WZW coupling
constant. Since, to give $c=26$, $k$ is small\foot{Rather than the
geometrically suggested $c \sim 2$, see \ref\gstr{S. Giddings and A.
Strominger, {\it Exact Black Fivebranes in Critical Superstring
Theory, } UCSBTH-91-35 Santa-Barbara preprint (1991).}\ for another
way to get $c=26$.}, it is lucky that the two dimensional nature of
the solution opens a better justification to space-time arguments.
They turn out to be exactly applicable for the ``tachyon'' states,
which dominate this solution \dvv. We will check the effects of
winding doubling in the \pfic\ conformal field theory and substantiate
them independently of space-time arguments (This check is encouraging
for bolder applications of space-time arguments).

  So far we were disucssing the Euclidean region \I\ formed gauging
the vectorial $U(1)$ subalgebra generated by $\sigma_2$. Let us see
how similar gauge conditions look through regions \III\ and \V. We
will get the Euclidean picture of these regions starting from the
Minkowski version. For region \I\  the later can be found writing
\eqn\eulm{ g_{\lower1pt\hbox{$\scriptstyle \,{\rm I},~mink.$}}=
e^{\half t_L \sigm_3} e^{\half r \, \sigm_1} e^{\half t_R \sigm_3}. }
and gauging the $U(1)$ generated by $\sigma_3$. The gauge where $g$
is symmetric gives the Minkowski line element $ds^2= dr^2 - \, \tanh^2
r \, d t^2$. The other regions follow in Kruskal coordinates $u,v$
\blw, where they can be put together. In region \I, $u= \half \sinh r
\, e^t,~ v= -\half \sinh r \, e^{-t}$. The line element is \eqn\krus{
ds^2= {du \, dv \over 1\,- \,uv}~~.} Region \I\  is $uv<0$, region
\III\ $0<uv<1$, bounded between the horizon $uv=0$ and the singularity
$uv=1$ and region \V\  is past the later. Kruskal description follows
from $SL(2,\IR)$ when parametrizing \eqn\vuab{g = \pmatrix {a & u \cr
-v &b },~~ ab + uv =1\,.} We continue this description to region \III,
where  $u= \half \sin r \, e^t,~ v= \half \sin r \, e^{-t}$ and $ds^2=
dr^2 - \, \tan^2 r \, d t^2$. $SL(2,\IR)$ is now parametrized
\eqn\eulrt{ g_{\lower1pt \hbox{$\scriptstyle \,{\rm III},~mink.$}}=
e^{\half t_L \sigm_3} e^{\i2 r \, \sigm_2} e^{\half t_R \sigm_3}. }

  Wick rotation in region \I\ turns \eulm\ into \eul, changing the
gauged $U(1)$ subgroup $e^{i \alpha \sigm_3}$ into $e^{ \alpha
\sigm_2}$ (acting vectorially as a similarity transformation) and the
line element into $ds^2= dr^2 + \, \tanh ^2 r \, d \theta^2$. In
region \III\ \eulrt\ chages to \eqn\eulrtm{ g_{\lower1pt
\hbox{$\scriptstyle \,{\rm  III},~euc.$}}= e^{\i2 \thet_L \sigm_3}
e^{\i2 r \, \sigm_2} e^{\i2 \thet_R \sigm_3} } ($\sigm_3$ cannot turn
into $\sigm_2$ already used by $r$) giving $g_{\lower1pt
\hbox{$\scriptstyle \,{\rm III},~euc.$}} \in SU(2)$! The line element
is $ds^2= dr^2 + \, \tan ^2 r \, d \theta^2$; as a check, near the
horizon $r=0$, it agrees with $ds^2$ for the sphere\foot{The
difference further on, shows that cosetting in current algebra does
not give the coset manifold, but is rather done by integrating out the
gauge field. The bilinear form in the gauge field $A$ is thus
responsible for the singularity, demonstrating the importance of
string effects in this context.} $S^2= SU(2)/U(1)$. $SU(2)_k/ U(1)_k$
in the Euclidean region \III, will lead us to \pfs\  \ref\zf{A.B.
Zamolodchikov and V.A. Fateev, \jep 62, 215, 1985. .} and their
duality, after further examination of the Euclidean solution and its
winding.

  As a further check for the Euclidean regions we have found and in
order to establish the way they are attached to each other, we will
find a path in $SL(2,\IR)$ through all of them. This is done by
imposing the condition\foot{providing a space-time description for
left movers on the world-sheet, which is sometimes interesting by
itself \witorb.} $\theta_R=0$, ($a^2 + v^2 = b^2 + u^2$ in the
parametrization of \vuab\ ) in addition to the gauge conditions. These
read $u+v=0$ ($g_{\hbox{$\scriptstyle \,{\rm I}$}}$ symmetric) in
region I, $(u+v)^2\,+\,(a+b)^2=4$ in \III\  ($g \in SL(2,\IR)$ always
satisfies $(u+v)^2\,+ \,(a+b)^2 \geq 4$, saturated at the horizon ${1
\, 0 \choose 0 \, 1 }$ and singularity ${{\phantom{ \scriptscriptstyle
-}0 \, 1~ \choose {\scriptscriptstyle -}1 \, 0~ }}$.) and $a+b=0$, in
\V\  (dual to $u+v=0$,). This path is parametrized by
\eqn\gees{\eqalign{ g_{\rho}=\pmatrix{ \cosh \rho & \sinh  \rho \cr
\sinh \rho &  \cosh  \rho \cr}& {\rm~~~~in~~I,~~~~~~~~~~~~}
g_{\omega}=\pmatrix {~\, \,\cos  \, \omega & \sin  \omega \cr -\sin
\omega &    \cos  \omega \cr}  {\rm~~in~~III~~}\cr  {\rm and~~}
g_{\sigma}= &\pmatrix{~\,\,\sinh \, \sigma & ~\, \,\cosh \,\sigma \cr
-\cosh \sigma & -\sinh  \sigma \cr} {\rm ~~in~~ V.}\cr}} $g_{\omega}$
belongs to $SU(2)$ as well (covering the intersection of the two
groups).

  Now we can look closely for winding in the Euclidean solution. It is
interestingly contrasted with a cylinder, which is the Euclidean
thermal version of flat Minkowski space \polch, whose set of winding
is topologically stable. For special values of the radius namely $k$
times the self-dual radius, they lead to $k+1$ conformal blocks\foot
{In addition to giving the primary operators, they also have effects,
like holomorphic factorization in the $\tau$ dependence of the torus
partition function, whose subtlety in our case we will see soon.}. The
Euclidean black hole is a bit different.  The cigar shaped region \I\
by itself, would drive us into the conclusion that winding is not
topologically stable here. This is not the case, since we have already
established that region \III\  is attached to region \I\  at $r=0$, by
threading a path between them.  Argued differently, the horizon, a
Minkowski light-cone in Kruskal coordinate is Wick rotated, giving
$r=0$, which is the point common to the regions. In fact, we are going
to argue that the winding is actually doubled, by closely studying the
\pfic\ theory in region \III\ and latter continuing it to the other
regions.

  The study of \pfic\ conformal field theories \zf\ was motivated by
critical two-dimensional $Z_k$ clock models in Statistical Mechanics
\ref\itz{ C. Itzykson, H. Saleur and J-B. Zuber, {\it Conformal
Invariance and Applications to Statistical Mechanics} (World
Scientific, 1988).}. They provide one of the best studied coset
models, $SU(2)_k/U(1)_k$. In addition to the role of a useful
laboratory for $U(1)$ cosetting they also happened to serve as the
centerpiece for the Euclidean string black hole solution. Most notable
in these models, is the wealth of observables ($\sim k^3$). We will
concentrate in this note on the set of order operators $\sigma_l$ and
the set of dis\-order operators $\mu_l$. The $\mu_l$'s are known to be
non-\-local with respect to the $\sigma_l$'s and correlation functions
containing both have cuts \ref\bpz{A.A. Belavin, A.M. Polyakov and
A.B. Zamolodchikov, \nub 241, 333, 1984. .}. $\sigma$'s and $\mu$'s
follow from gauging different $U(1)$ subalgebras of $SU(2)_k$, the
vectorial (electric) $U(1)$ gives the  $\sigma$'s and the axial
(magnetic) $U(1)$ gives the $\mu$'s. In other words, they result from
quantizing different sets of zero modes (windings) in the theory
thereby demonstrating doubling.

  \def\nord{{\rm :}}
  \def\phib{\overline{\phi}}
  \def\dl{\partial}

  To see that this is indeed the case we note that $\sigma$'s and
$\mu$'s are related by Kramers-Wannier duality \ref\dual{N.A. Kramers
and G.H Wannier, \pro 60, 252, 1941. .}. This duality, which relates
magnetic cosetting to electric cosetting, works the same way as the
duality found in the string black hole solution \amit\ \dvv. For more
careful examination, we observe that the cut structure in the \pfic\
amplitudes is the inverse of the cut structure between electric and
magnetic $U(1)$ vertices. In chapter 5 of \zf\  the $SU(2)_k$
holomorphic current $J^3(z) = \dl_z \phi(z)$, is taken to be the
holomorphic $U(1)$ current, $\phi(z)$ being the holomorphic part of
the scalar field $\Phi(z,\zb)$. By formula (5.10) in \zf\ \eqn\sigl{
\sigma_l(z,\zb) \,\, \nord \exp \Bigl({i\, l \over \sqrt{k} }
\bigl(\phi (z)+ \phib (\zb) \bigr) \Bigr) \nord = \sigma_l(z,\zb)\, \,
\nord \exp \Bigl({i\, l \over \sqrt{k}} \Phi (z,\zb) \Bigr) \nord }
gives a cutless $SU(2)_k$ correlator with \eqn\mun{\mu_n(z,\zb)\, \,
\nord \exp \Bigl({i\, n \over \sqrt{k}} \bigl(\phi (z)- \phib (\zb)
\bigr) \Bigr) \nord = \mu_n(z,\zb) \,\,\nord \exp \Bigl({i\, n \over
\sqrt{k}} \tilde {\Phi} (z,\zb) \Bigr) \nord ~~.} Thus, all the \pfic\
$Z_k$ cut structure is between the electric $\Phi(z,\zb)$ and magnetic
$\tilde{\Phi}(z,\zb) =\phi(z) - \phib(\zb)$ vertices. A bit more
algebraic version of the description suggested here for $\sigma$ and
$\mu$ is found introducing the rational torus $U(1)_k\subset SU(2)_k$
by which we coset each time. Then $\sigma$ are characterized as the
operators which are local with respect to $J^3(z)$ and $J^+(z)^k \,
\bar{J}^+(\zb)^k$. $\mu$ are rather local with respect to $J^+(z)^k \,
\bar{J}^-(\zb)^k$. Doubling of winding is noted realizing that
different sets of winding give the $\sigma$'s and the $\mu$'s upon
quantization, since they came about, cosetting different $U(1)$'s. All
the operators in the \pfic\ theory will be discussed in this laguage
elsewhere.

  The cut structure in \pfs\ provides an important property shared
with non-critical strings, or Liouville theory and hence an additional
relationship \ref\mash{E. Martinec and S. Shatashvili, {\it Black Hole
Physics and Liouville Theory,} Chicago preprint, EFI-91-22.} between
the black hole solution and non-critical strings. If we consider the
$\sigma$'s as operators in the bulk of the system, then the $\mu$'s
seem to introduce boundaries in the form of cut lines between the
various $\mu$'s. It is also natural to specify modified boundary
conditions on the cylider (or torus) as though a cut is running
between two $\mu$'s on the boundary\ref\carn{J. Cardy, \nubf 275, 17,
200, 1986. .}\  \ref\cardl{J. Cardy, {\it Conformal Invariance and
Statistical Mechanics,} Les Houches Lectures (1988).}. This results in
more modular invariant combinations, under a subgroup of the mapping
class group\foot{Along with the ambiguity in specifying the chiral
algebra and the cuts in the amplitudes, this makes the definition of
conformal blocks tricky. For example, the Ising model has 3 blocks
under the Virasoro algebra but only 2 blocks when the chiral algebra
consists of a free Majorana fermion. In this case, the modular
invariance is under the group generated by $S$ and $T^2$. These
subtleties in defining conformal blocks are typical to doubling of
winding.}. This distinction between operators, naturally specified on
points and operators associated with boundaries was made by Seiberg
and Moore\ref\naliu{N. Seiberg, {\it Notes on Quantum Liouville Theory
and Quantum Gravity,} Lectures at the 1990 Yukawa International
Seminar, Common Trends in Mathematics and Quantum Field Theory,
Rutgers preprint RU-90-29.}\ \ref\mus {G. Moore and N. Seiberg, {\it
{}From Loops to Fields in 2-d Quantum Gravity.} Rutgers preprint,
RU-91-29.}\ (the later were associated with normalizable wave
functions) in the context of two dimensional gravity. In our case
operators and states are dual.

  Further understanding of the \pfic\ theories as coset model is
required in order to get semiclassical intuition. Large $k$ is
currently under investigation and seems hopeful, since  self dual
$Z_k$ clock models fall withing a Kosterlitz-Thouless phase for $k>4$
and should not be too sensitive to $1/k$ corrections (as well as to
moving off criticallity, possibly by Ginzburg-Landau formulation. A
further speculation will be that lattice models could do as well as
their coninuum limits, in accord with discrete or topological
approaches to 2-d gravity.). It should also be mentioned that \pfic\
models continues smoothly accross the horizon to region \I\ with
$\Phi$ as $\theta$. Instead of warrying about continuation across the
singularity, we will simply use the \pfic\ duality to see that
$\bar{\Phi}$ continues to $\tilde{\theta}$, further empasizing
that Kramers-Wannier duality is equivalent to the duality found in the
black hole solution\amit\ \dvv.

  The importance of winding for string thermodynamics was already
mentioned above. $\ell\,$fold Euclidean winding states becoming
massless, was also interpreted in string thermodynamics \ref\hag{M.
Spiegelglas, \plb 220, 391, 1989. .}. Much the same way that by
becoming massless, a minimal winding state signifies that higher
string excitations start to dominate the string free energy; these
$\ell\,$fold windings become massless when highly excited $\ell$
identical strings states get to dominate the ensemble of $\ell$
identical string excitations. This interpretation was shown to be
consistent with Bose-Einstein as well as Fermi-Dirac quantum
statistics, in the bosonic and fermionic string theories. It is
tempting to abstract from this theromdynamical argument that the
Euclidean winding states are related to quantum correlations between
identical particles. By that we would learn that identical particles
correlations could behave differently in the black hole case. One
could speculate that the doubling of winding states, would naively
present itself as a $Z_2$ additional quantum number (``color''). Along
with implications to black-hole thermodynamics, this  should be
examined more carefully.

  A calculation of the partition function on the torus or other
amplitudes will be very helpful to clarify this issues as well as the
physical spectrum. It will also be interesting to relate the arguments
found here to the extra twisted states found at \ref\bo{M. Henningson,
S. Hwang, P.Roberts and B. Sundborg, {\it Modular Invariance of
SU(1,1) Strings}, G\"{o}teborg and Stockholm preprint.} in the
$SL(2,\IR)$ theory.

\smallskip

  {\bf Acknowledgments:} I am grateful to J. Avron, M. Dine, F.
Englert, Y. Feinberg, M. Marinov,  B. Reznik, N. Rosen, J.
Sonnenschein and S. Yankielowicz for valuable discussions. S.
Yankielowicz has convinced me how tricky the no-ghost situation is and
cosequenly focused my interest on the black hole solution. This work
was supported in part by US Israel Binational Science Foundation
(BSF), by the Israeli Academy of Sciences and by the Technion V.P.R.
Fund.

\bigskip

\listrefs
\bye